\documentclass[twocolumn,nofootinbib]{revtex4-1}
\usepackage{latexsym,epsfig,amssymb, amsmath,nicefrac}
\usepackage{epsfig,graphicx}
\usepackage[usenames, dvipsnames]{color}
 
\newcommand{\vt}{\vartheta}

              \newcommand{\rf}[1]{(\ref{#1})}

\def\bfone{\relax{\rm 1\kern-.35em 1}}

\newcommand{\be}{\begin{equation}}
\newcommand{\ee}{\end{equation}}
\newcommand{\ben}{\begin{displaymath}}
\newcommand{\een}{\end{displaymath}}
\newcommand{\bea}{\begin{eqnarray}}
\newcommand{\eea}{\end{eqnarray}}

\newcommand{\bean}{\begin{eqnarray*}}
\newcommand{\eean}{\end{eqnarray*}}

\newcommand{\vp}{\varphi}

\def\Kahler{K\"{a}hler~}

\def\K{K{\"a}hler}

\makeatletter \@addtoreset{equation}{section} \makeatother

\begin{document}

\title{\Large{Cosmological Attractors and Asymptotic Freedom of the Inflaton Field}}

\author{Renata Kallosh and Andrei Linde}

\affiliation{Department of Physics and SITP, Stanford University, \\ 
Stanford, California 94305 USA}

\begin{abstract}
We show that the inflaton coupling to all other fields is exponentially suppressed during inflation in the  cosmological $\alpha$-attractor models. In the context of supergravity, this feature is a consequence of the underlying hyperbolic geometry of the moduli space  which has a flat direction corresponding to the inflaton field. A combination of these factors protects the asymptotic flatness of the inflaton potential. 
 
\end{abstract}

\maketitle

\smallskip


\section{Introduction}\label{intro}

Recent observational data  \cite{Planck:2013jfk} attracted lots of attention to large field inflationary models with plateau-like inflaton potentials. 
The plateau potential of the required type for the first time was introduced in the  chaotic inflation model in supergravity \cite{Goncharov:1983mw}. At large  values of the inflaton field $\varphi$, the inflaton potential in that model was 
\be\label{app}
V(\phi) = V_{0}  - {8V_{0} \over 3}\, e^{-\sqrt 6  |\vp| }  \ .
\ee 
The Starobinsky model originally was based on investigation  of the conformal anomaly in quantum gravity  \cite{Starobinsky:1980te}. Then it was re-formulated as a  theory $f(R) =R+c R^{2}$   \cite{Barrow:1983rx,Starobinsky:1983zz}. Few years later, it was found that one can further re-formulate it as a theory of a scalar field with a potential
\be\label{star}
V = V_{0}\left(1- e^{-\sqrt{2/3}\,\varphi}\right)^2 \approx V_{0}  - 2 V_{0}\, e^{-\sqrt{2/3}\,\varphi} \ ,
\ee
where the last part of this equation is valid during inflation  at $\varphi \gg 1$ \cite{whitt}. The  coefficients ${8V_{0} \over 3}$ in \rf{app} and $2 V_{0}$ in front of the exponent in \rf{star} are not very important since they can be absorbed in  redefinition of the inflaton field $\varphi$. The potential of the Higgs inflation 
model for $\varphi > 0$ in the original version of this theory has the same asymptotic form \cite{Salopek:1988qh,Bezrukov:2007ep}.   

However, for a while it was not quite clear whether there is any unifying principle which would explain why these very different theories have very similar potentials and nearly coinciding predictions $n_{s} = 1-2/N$ and $r = O(N^{{-2}})$, successfully matching the observations   \cite{Planck:2013jfk}.
In general, no symmetry principles prevent  one from generalizing the function $f(R)$ in the Starobinsky model in  a way that would make the inflationary plateau \rf{star} very short. Strong SUSY breaking could modify the potential of the GL model \cite{Goncharov:1983mw} at large $\vp$ \cite{Linde:2014hfa}.
Quantum corrections  may preserve some features of the Higgs inflation \cite{Barvinsky:2009fy,Ferrara:2010in,Bezrukov:2010jz}, and yet they may significantly modify the shape of the inflationary potential and its observational predictions, see e.g. Figs. 1 and 5 in \cite{Bezrukov:2014bra}. Moreover, the original choice of the non-minimal coupling and of the inflationary potential required in \cite{Salopek:1988qh,Bezrukov:2007ep} was very special. This choice was significantly generalized in \cite{Kallosh:2013tua}, but it still required an unexplained correlation between the functional form of the non-minimal coupling of the Higgs field to gravity and its potential. 

These issues are not particularly important if our only goal were to build  advanced inflationary models  matching all available cosmological data. There are methods that allow one to construct  inflationary models in supergravity with an arbitrary inflationary potential, which can describe any combination of inflationary parameters  $n_{s}$ and $r$ \cite{Kawasaki:2000yn,Kallosh:2010ug,Ferrara:2013rsa}. Using these and other approaches one can implement the Starobinsky model in supergravity \cite{Cecotti:1987sa,Kallosh:2013xya,Farakos:2013cqa,Ferrara:2013rsa}, or find the supergravity-based models with more general plateau potentials $V_{0} (1- e^{-\sqrt{2/3\alpha}\,\varphi})^2$\, \cite{Ferrara:2013rsa}. But since these methods  can describe {\it arbitrary}\, inflationary potentials, they leave unexplained the success of the plateau potentials and the mysterious coincidence of the results obtained in different models \cite{Starobinsky:1980te,Goncharov:1983mw,Salopek:1988qh,Bezrukov:2007ep}. Similarly, one may  reproduce some part of a plateau potential in the context of the no-scale supergravity \cite{Ellis:2013xoa}, or string theory \cite{Cicoli:2008gp,Burgess:2016owb}. One can successfully describe observational data using these models, but it requires fine-tuning, and the resulting inflationary potentials typically blow up at large $\vp$, see e.g. Fig. 1 in  \cite{Ellis:2013xoa} and Fig. 1 in \cite{Burgess:2016owb}.

Fortunately, one can incorporate most of the existing  inflationary models with plateau potentials, including the GL model   \cite{Goncharov:1983mw}, the Starobinsky model  \cite{Starobinsky:1980te,Barrow:1983rx,Starobinsky:1983zz} and some generalized versions of the Higgs inflation \cite{Salopek:1988qh,Bezrukov:2007ep,Kallosh:2013tua} into a broad class of models  called $\alpha$-attractors  \cite{Kallosh:2013xya,Kallosh:2013hoa,Kallosh:2013yoa,Cecotti:2014ipa}. As we will review in section \ref{basic}, the flatness of the inflaton potential of the $\alpha$-attractors is explained and protected by the existence of a pole in the kinetic term of the scalar fields $\phi$ in such models \cite{Galante:2014ifa}. Upon transformation to the canonical field variables $\vp$, any non-singular inflaton potential $V(\phi)$ acquires a universal plateau-like form at large values of the inflaton field $\vp$ \cite{Kallosh:2013xya,Kallosh:2013hoa,Kallosh:2013yoa,Cecotti:2014ipa,Galante:2014ifa}.

In the supergravity context, the universal properties of these theories can be traced back to the hyperbolic geometry of the moduli space and the flatness of the \K\ potential in the inflaton direction  \cite{Kallosh:2015zsa,Carrasco:2015uma,Carrasco:2015rva,Kallosh:2016ndd}. All of these models make very similar cosmological predictions providing good match to the latest cosmological data. Advanced versions of these models can successfully describe not only inflation, but also the theory of dark energy/cosmological constant and supersymmetry breaking \cite{Ferrara:2014kva,Kallosh:2015lwa,Carrasco:2015pla,Carrasco:2015iij}, while preserving the asymptotic flatness of the inflaton potential.  

In this paper we will show that these theories have yet another interesting property: The canonically normalized  inflaton field $\vp$ in the broad class of $\alpha$-attractor models couples to other fields with coupling constants which become {\it exponentially small} for large values of the inflaton field $\varphi \gg 1$. Masses of all particles interacting with this field at its large values exponentially fast approach some constant values. These features have the same geometric interpretation as the asymptotic flatness of the inflaton potential in these models. In section \ref{simplemodel} we  illustrate this statement in a simple $\alpha$-attractor  model describing an inflaton coupled to matter, and then we generalize our results to supergravity in section \ref{sugra}. 
Finally, in section \ref{CWsection} we will show that the asymptotic flatness of the inflaton potential is preserved with an account of the one-loop quantum corrections to the inflaton potential.

\section{The basic $\alpha$-attractor model }\label{basic}
On a purely phenomenological level, the main features of inflation in all of these models can be represented in terms of a single-field toy model with the Lagrangian  \cite{Kallosh:2013hoa,Kallosh:2013yoa,Cecotti:2014ipa,Galante:2014ifa,Kallosh:2015zsa}
 \be
 {1\over \sqrt{-g}} \mathcal{L} = { R\over 2}   -  {(\partial_{\mu} \phi)^2\over 2(1-{\phi^{2}\over 6\alpha})^{2}} - V(\phi)   \,  .
\label{cosmo}\ee
Here $\phi(x)$ is the scalar field, the inflaton.  The origin of the pole in the kinetic term will be explained in the context of hyperbolic geometry in supergravity and string theory in Sec. \ref{sugra}.
The parameter  $\alpha$  can take any positive value. In the limit $\alpha \rightarrow \infty$ this model coincides with the standard chaotic inflation  with a canonically normalized field $\phi$ and the inflaton potential $V(\phi)$  \cite{Linde:1983gd}. 

However, for any finite values of $\alpha$, the field $\phi$ in \rf{cosmo} is not canonically normalized. It must satisfy the condition $\phi^2<6\alpha$, for the sign of the inflaton kinetic term to remain positive.  We assume that the theory is well defined for $\phi^2<6\alpha$, in particular,  the potential $V(\phi)$ and its derivatives are non-singular for $\phi^2<6\alpha$. 
 One can easily go to canonically normalized variables $\vp$ by solving the equation ${\partial \phi\over 1-{\phi^{2}\over 6\alpha}} = \partial\vp$, which yields
\be\label{tanh} 
\phi = \sqrt {6 \alpha}\, \tanh{\varphi\over\sqrt {6 \alpha}} \ .
\ee
The full theory, in terms of the canonical variables, becomes
 \be
 {1\over \sqrt{-g}} \mathcal{L} = { R\over 2}   -  {(\partial_{\mu}\varphi)^{2} \over 2}  - V\big(\sqrt {6 \alpha}\, \tanh{\varphi\over\sqrt {6 \alpha}}\big)   \,  .
\label{cosmoqq}\ee
Note that in the limit $\phi \to 0$ the variables $\phi$ and $\varphi$ coincide; the main difference appears in the limit $\phi \to \sqrt{6 \alpha}$: In terms of the new variables, a tiny vicinity of the boundary of the moduli space at $\phi=\sqrt{6\alpha}$ stretches and extends to infinitely large $\varphi$. As a result, generic potentials $V(\phi) = V(\sqrt {6 \alpha}\, \tanh{\varphi\over\sqrt {6 \alpha}})$ at large $\vp$ approach an infinitely long dS inflationary plateau with the height corresponding to the value of $V(\phi)$ at the boundary:
\be
V_0 = V(\phi)|_{\phi = \pm \sqrt {6 \alpha}} \ .
\ee

To understand what is going on in this theory, let us consider, for definiteness, positive values of $\phi$ and study a small vicinity of the point $\phi = \sqrt {6 \alpha}$, which  becomes stretched to infinitely large values of the canonical field $\vp$  upon the change of variables $\phi \to \vp$. If the potential $V(\phi)$ is non-singular at the boundary  $\phi = \sqrt {6 \alpha}$, we can expand it in series with respect to the distance from the boundary:
\be
V(\phi) = V_{0} + (\phi-\sqrt {6 \alpha})\, V'_{0} +O(\phi-\sqrt {6 \alpha})^{2} \ .
\ee
where we denote $V'_{0} = \partial_{\phi}V |_{\phi = \sqrt {6 \alpha}}$. 

In the vicinity of the boundary $\phi=\sqrt {6 \alpha}$, the relation \rf{tanh} between the original field variable $\phi$ and the canonically normalized inflaton field $\vp$ is given by
\be\label{tanh2} 
\phi = \sqrt {6 \alpha}\, \left(1 - 2 e^{-\sqrt{2\over 3\alpha} \varphi }\right)\ ,
\ee
up to the higher order terms $O\bigl(e^{-2\sqrt{2\over 3\alpha} \varphi }\bigr) $. At $\vp \gg \sqrt \alpha$, these  terms are exponentially small as compared to the terms $\sim  e^{-\sqrt{2\over 3\alpha} \varphi }$, and the potential acquires the following asymptotic form:
\be\label{plateau}
V(\vp) = V_{0} - 2  \sqrt{6\alpha}\, V'_{0}\ e^{-\sqrt{2\over 3\alpha} \varphi } \ .
\ee
Note that the constant $2  \sqrt{6\alpha}\, V'_{0}$ in this expression can be absorbed into a redefinition (shift) of the field $\vp$. This implies that if inflation occurs at large $ \vp \gg \sqrt{\alpha}$, all inflationary predictions in this class of models of the potential $V(\phi)$ are determined only by the value of the potential $V_{0}$ at the boundary and the constant $\alpha$. For $\alpha = O(1)$, the  amplitude of inflationary perturbations, the prediction for the spectral index $n_{s}$ as well as the tensor to scalar ratio $r$ match observational data under a single nearly model-independent condition 
\be
{V_{0}\over  \alpha} \sim 10^{{-10}} \ .
\ee

These results were explained in  \cite{Kallosh:2013hoa,Kallosh:2013yoa,Kallosh:2015lwa} and formulated in a particularly general way in \cite{Galante:2014ifa}:  The kinetic term in this class of models has a pole at the boundary of the moduli space. If inflation occurs in a vicinity of such a pole, and the potential near the pole has a finite and positive first derivative, all other details of the potential and of the kinetic term far away from the pole (from the boundary of the moduli space) become unimportant for making cosmological predictions. In particular, the spectral index depends solely on the order of the pole, and the tensor-to-scalar ratio relies on the residue  \cite{Galante:2014ifa}. All the rest is practically irrelevant, as long as the field after inflation falls into a stable minimum of the potential, with a tiny value of the vacuum energy, and stays there.
Stability of the inflationary predictions with respect to even very strong modifications of the shape of the potential outside a small vicinity of the boundary of the moduli space is the reason why these models are called cosmological attractors.

This new class of models accomplishes for inflationary theory something similar to what inflation does for cosmology. Inflation stretches the universe making it flat and homogeneous, and the structure of the observable part of the universe becomes very stable with respect to the choice of initial conditions in the early universe. Similarly, stretching of the moduli space near its boundary upon transition to canonical variables makes inflationary potentials very flat and results in predictions which are very stable with respect to the choice of the inflaton potential.

This addresses the often presented argument that the shape of the inflationary potential in large-field inflation and its cosmological predictions must be unstable with respect to higher order corrections to the potential at super-Planckian values of the field. In the new class of models, the range of the original field variables $\phi$ for $\alpha \lesssim 1$ is sub-Planckian, and the shape of the potential in terms of the canonical inflaton field is very stable with respect to the choice of the original potential \rf{plateau}, all the way to infinitely large values of the inflaton field. At large values of the inflaton field, its potential, as well  as the potentials of all other fields, have shift symmetry with respect to the inflaton field. This fact helps to solve the problem of initial conditions in such theories  \cite{Carrasco:2015rva,East:2015ggf}.

One may wonder whether quantum corrections can change the situation. For example, if the inflaton field gives mass $g\phi$ to some other fields, and the value of the field $\phi$ can be arbitrarily large, then the masses of these additional fields at super-Planckian values of the inflaton field become extremely large, which could lead to  large quantum corrections altering the shape of the original potential  \cite{Coleman:1973jx,Linde:1975sw,Weinberg:1976pe,Krive:1976sg,Linde:2005ht}.

\section{A simple $\alpha$-attractor model including matter fields}\label{simplemodel}

The basic reason  can be explained by  adding to our toy model (\ref{cosmo}) the second field $\sigma$:
 \be
 {1\over \sqrt{-g}} \mathcal{L} = { R\over 2}   -  {(\partial_{\mu} \phi)^2\over 2(1-{\phi^{2}\over 6\alpha})^{2}} -  {(\partial_{\mu} \sigma)^2\over 2}  - {m^{2}\over 2} \phi^{2}   - {g^{2}\over 2}\phi^{2}\sigma^{2} - {M^{2}\over 2} \sigma^{2} .
\label{cosmo2}\ee
In terms of canonical fields $\vp$ with the kinetic term  ${\partial_{\mu} \vp^2\over 2}$, the  potential is 
 \be
 V(\varphi,\sigma) =    3\alpha (m^{2 }+g^{2}\sigma^{2})  \, \tanh^{2}{\varphi\over\sqrt {6 \alpha}} +{M^{2}\over 2} \sigma^{2} .
\label{cosmo3}\ee
The potential depends on $|\vp|$. During inflation at  $|\varphi | \gg \sqrt\alpha$, one can use the asymptotic equation
\be
\tanh^{2}{|\varphi |\over\sqrt {6 \alpha}} = 1-4 e^{-\sqrt{2\over 3\alpha}|\varphi |} + O(e^{-2\sqrt{2\over 3\alpha}|\varphi |}) \ .
\ee
For notational simplicity, we will study positive values of $\vp$. The potential at  $\varphi \gg \sqrt\alpha$ is equal to 
\be
 V(\varphi,\sigma)=  3\alpha (m^{2 }+g^{2}\sigma^{2})  \, (1-4 e^{-\sqrt{2\over 3\alpha}\varphi})  +{M^{2}\over 2} \sigma^{2} , 
 \ee 
 up to exponentially small higher order terms $3\alpha (m^{2 }+g^{2}\sigma^{2}) O(e^{-2\sqrt{2\over 3\alpha}\varphi})$. The potential $ V(\varphi,\sigma)$ has a minimum with respect to $\sigma$ at $\sigma = 0$. The  inflaton potential at $\sigma = 0$ and  large $\varphi$ is  
 \be\label{ppotcan}
 V(\vp) = 3\alpha  m^{2 } \tanh^{2}{\varphi\over\sqrt {6 \alpha}} \ . 
  \ee
During inflation at $\vp \gg \alpha$, this potential is
\be\label{ppot}
 V(\vp) \approx 3\alpha\, m^{2 }  \, (1-4 e^{-\sqrt{2\over 3\alpha}\varphi})\ . 
 \ee
Mass squared of the canonically normalized field $\vp$ is given by the second derivative of $ 3\alpha  m^{2 } \tanh^{2}{\varphi\over\sqrt {6 \alpha}}$. At $\vp\ll \sqrt\alpha$  one has
\be
m^{2}_{\vp} = m^{2} \ ,
\ee
but at $\vp\gg \sqrt\alpha$ one finds 
\be\label{mp}
m^{2}_{\vp} = -8 m^{2} e^{-\sqrt{2\over 3\alpha}\varphi} \ .
\ee
The mass of the field $\sigma$ is equal to    
\be\label{ms}
m^{2}_{\sigma} = M^{2} +  g^{2}\phi^{2} =   M^{2} + 6\alpha g^{2}(1-4 e^{-\sqrt{2\over 3\alpha}\varphi}) \ .
\ee
where the last relation is valid for at $\vp\gg \sqrt\alpha$.

The strength of interactions of  the  inflaton field with itself and with the field $\sigma$ during inflation at $\sigma = 0$ can be described in terms of the coupling constants of  canonically normalized fields, such as $\lambda_{\varphi,\varphi,\varphi,\varphi} = \partial^{4}_{\varphi}  V(\varphi, \sigma)_{{|}_{\sigma = 0}}$ or $\lambda_{\varphi,\varphi,\sigma,\sigma} = \partial^{2}_{\varphi} \partial^{2}_{\sigma} V(\varphi, \sigma)_{{|}_{\sigma = 0}}$. As one can easily see, all such couplings are suppressed by the exponentially small coefficient $e^{-\sqrt{2\over 3\alpha} \varphi }$. 

That is why we called the inflaton field ``asymptotically free''. By that, we  mean the {\it exponentially small} strength of interactions of the field $\vp$ with all other fields  at large $\vp$, rather than the {\it logarithmically small} strength of interactions at large momenta, as in QCD.  Similar results have been obtained for the Higgs inflation \cite{Bezrukov:2010jz}, but there they apply only for a  specific choice of the inflaton potential, see a discussion in the Introduction. Meanwhile our conclusions apply to the models with any potential $V(\phi,\sigma)$ as long as this potential and its derivatives are non-singular at the boundary $\phi = \sqrt{6\alpha}$.

Indeed, consider any potential $V(\phi,\sigma)$.  The large $\varphi$ limit corresponds to $\phi \to \sqrt{6\alpha}$. In this limit 
\be 
\partial_{\varphi} V(\phi, \sigma) = {\partial \phi \over \partial \varphi }\  \partial_{\phi} V(\phi, \sigma)\ , 
\ee
where  $\partial_{\phi} V(\phi, \sigma)$  should be calculated near the boundary, in the limit $\phi \to \sqrt{6\alpha}$. At large $\vp$, the derivative ${\partial \phi \over \partial \varphi }$   is given by $2 \sqrt{2\over 3\alpha }\, e^{-\sqrt{2\over 3\alpha}\varphi}$, so  one finds
\be 
\partial_{\varphi} V(\phi, \sigma)= 2 \sqrt{2\over 3\alpha }\, e^{-\sqrt{2\over 3\alpha}\varphi}\  \partial_{\phi} V(\phi, \sigma)_{|_{\phi \to \sqrt{6\alpha}}} \ .
\ee
As long as $V(\phi, \sigma)$ and its derivatives are non-singular at the boundary $\phi = \sqrt{6\alpha}$, one finds that $\partial_{\varphi} V(\phi, \sigma)$ is exponentially suppressed by the factor $e^{-\sqrt{2\over 3\alpha}\varphi}$. All higher derivatives with respect to $\varphi$ and $\sigma$, describing strength of interactions of these fields with each other, are suppressed by the same exponentially small factor, for example
\be 
 \lambda_{\varphi, \sigma,\sigma} = \partial_{\varphi}\partial^{2}_{\sigma} V(\phi, \sigma)= 2 \sqrt{2\over 3\alpha }\, e^{-\sqrt{2\over 3\alpha}\varphi}\ \partial_{\phi}\partial^{2}_{\sigma} V(\phi, \sigma)_{|_{\phi \to \sqrt{6\alpha}}}\ .
\ee

\section{Hyperbolic geometry and plateau potentials of the inflaton field}\label{sugra}

We are turning now to supergravity versions of $\alpha$-attractors. There were many attempts to build successful inflationary models in supergravity. It was a rather difficult task mostly because of the presence of the term $e^{K}$ in the expression for the F-term potential. For the simplest choice of the \K\ potential $K = \Phi\overline\Phi$, the inflaton potential at large $\Phi$ was too steep, growing as $e^{\Phi\overline\Phi}$. One could compensate for this growth by a smart choice of a superpotential. This is what was achieved in the first model of chaotic inflation in supergravity, which has a plateau potential \rf{app} \cite{Goncharov:1983mw}. However, the real breakthrough happened almost two decades later, with the systematic use of the \K\ potentials with a flat direction corresponding to the inflaton field \cite{Kawasaki:2000yn,Kallosh:2010ug}. Even the unexpected early success in building an inflationary model with a plateau potential \cite{Goncharov:1983mw} was fully understood only much later, when it was realized that this model can be formulated as a model with a  \K\ potential with a flat direction \cite{Linde:2014hfa,Kallosh:2015lwa}.

The simplest example of such \K\ potentials is given by $K = (\Phi - \overline\Phi)^{2}/2 + S\overline S$, where $\Phi = (\vp +i \vartheta)/\sqrt 2$. Here $\vp$ is a canonical inflaton field. By taking a superpotential 
\be\label{KLR}
W = S f(\Phi)
\ee
 one finds a general family of the inflaton potentials \cite{Kallosh:2010ug}
\be
V = |f^{2}(\vp/\sqrt 2 )| \ .
\ee 
Dangerous terms like $e^{\Phi\overline\Phi}$ do not appear in the inflaton direction because $K = 0$ for $\Phi = \overline\Phi$.

This class of theories is very general; it can easily incorporate the models with the potential exactly coinciding with the Starobinsky-Whitt potential \rf{star}, as well as the T-models with the potential \rf{tanh}. However, in this approach the kinetic term of the inflaton field is canonical from the very beginning, and potentials $V$ with an infinite plateau require extreme fine-tuning of the function $f(\Phi)$. 

 The  \K\, potentials which are motivated by string theory and extended supergravity with ${\cal N}\geq 2$ supersymmetry have a logarithmic dependence on moduli and an associated $SL(2,\mathbb{R})$ symmetry, Poincar\'e disk geometry. Quantum corrections might break this symmetry to a discreet one, $SL(2,\mathbb{Z})$. However, the \K\, potential still requires a logarithmic dependence on moduli which is protected by this modular invariance. Such a logarithmic dependence always leads to poles in the kinetic terms for scalars. For example with $  \log (T+\overline T)$ we find the kinetic term ${\partial T \partial \bar T\over (T+\bar T)^2}$, and for $  \log (1-Z\bar Z)$ we find ${\partial Z \partial \bar Z \over (1-Z\bar Z)^2}$. 

Therefore to recover the advantages of the theories with the pole in the kinetic term, one may start with the theories with logarithmic \K\ potentials, such as $-3  \log (T+\overline T)$ in half-plane variables $T$ \cite{Cecotti:1987sa,Kallosh:2013xya,Cecotti:2014ipa,Kallosh:2015zsa} or $-3   \log ( 1 - Z \overline Z )$ in disk variables \cite{Kallosh:2013yoa}. For example, the theory with the superpotential $W = S f(T)$ \rf{KLR} with $f(T)  = 3M (T-1)$ and the \K\ potential
\be\label{cec}
K = -3 \log \left[T+\overline T - S\overline S + c (S\overline S)^{2}\right]
\ee
exactly reproduces the Starobinsky-Whitt plateau potential with $V_{0} = 3M^{2}/4$ \cite{Cecotti:1987sa,Kallosh:2013xya}, for $T =(\vp +i \vartheta)/\sqrt 2$. (The term $c (S\overline S)^{2}$ was required for stabilization of the field $S$ near $S = 0$.) 

A further progress in constructing $\alpha$-attractors in supergravity was achieved very recently when the transition was made from the \K\ potentials such as \rf{cec}, to their equivalent shift-symmetric counterparts, such as \cite{Carrasco:2015uma,Carrasco:2015pla,Carrasco:2015rva,Carrasco:2015iij,Kallosh:2016ndd},
\be \label{new-K}
 K = -{3\alpha\over 2} \log \left[{(T+\overline T)^2\over 4 T \overline T }\right] +S\overline S\, .
\ee
The new \K\ potentials are related  by a  \Kahler transformation to the original ones. However, the new \K\ potentials have a symmetry under the shift of the inflaton, which corresponds to the real direction $T+\overline T$, accompanied by the rescaling of the inflaton partner $T-\overline T$. During inflation, $T = \overline  T$ and therefore $K=0$,  which is obviously invariant under the inflaton shift  \cite{Carrasco:2015uma,Carrasco:2015pla,Carrasco:2015rva}. One can also formulate the required property of the \K\ potential as \cite{Kallosh:2016ndd}
\be
\partial_{T}K_{|_{T= \overline T}} = 0 \ .
\ee
The shift symmetry of the inflaton potential is only slightly broken by the superpotential.

In this class of models, the pole of the kinetic term occurs at $T = 0$. Any superpotential $W = Sf(T)$ with $f(T)$ non-singular at $T = 0$ leads to a non-singular potential $V(T) = |f^{2}(T)|$. Just like in our single-field models considered in sections \ref{basic}, \ref{simplemodel}, any real holomorphic function $f(T)$ with an absolute value growing towards $T = 0$ leads to a  plateau potential with respect to the canonically normalized inflaton field $\varphi$ such that $T = e^{-\sqrt{2/3}\,\varphi}$. It reproduces the Starobinsky potential for $f(T) \sim 1-T$. In this approach, the flatness of the potential is {\it not} affected by any non-singular corrections $\Delta f(T)$.

Similarly, instead of the \K\, potential  $-3 \alpha \log ( 1 - Z \overline Z )+S\overline S$ in disk variables \cite{Kallosh:2013yoa} it is convenient  to use an improved   \K\ potential for $\alpha$-attractors, which vanishes for real values of the field $Z$, which is the inflaton direction \cite{Carrasco:2015uma,Carrasco:2015pla,Carrasco:2015rva,Carrasco:2015iij,Kallosh:2016ndd}: 
\be\label{NewD}
K= -{3  \alpha\over 2}  \log  \left[ {( 1- Z \overline{Z} )^2  \over  (1-Z^2) (1- \overline{Z}^2)}  \right] + S \overline{S} \, .  
\ee
The two descriptions, in terms of the half-plane variables and disk variables, are directly related to each other  \cite{Cecotti:2014ipa,Kallosh:2015zsa}. In particular, half-plain variables $T$ are related to the disk variables $Z$ by the Cayley transform,
\be
Z= {T-1\over T+1}\, ,    \qquad T= {1+Z\over 1-Z} \, .
\label{Z1}
\ee 
All effects to be discussed below can be described using these and other \K\ potentials having similar properties. 

\section{Asymptotically free inflaton in a simple supergravity model}\label{sugra2}
We will use $Z$ variables and the \K\ potential \rf{NewD}, which will help us to more directly relate this theory to the toy models discussed iin sections \ref{basic}, \ref{simplemodel}.  
We will  consider a simplest model with the superpotential 
\be\label{sup}
W = \sqrt{3\alpha}\, m S Z \ .
\ee
One can represent $Z$ as
\be
Z =  {\phi+i z\over \sqrt{6\alpha}}\ . 
\ee
The kinetic term of the field $\phi$ is the same as in \rf{cosmo2}. As before, one can replace the field $\phi$  by a canonically normalized inflaton field $\phi =  \sqrt {6 \alpha}\tanh{\varphi\over\sqrt {6 \alpha}}$, see eq. \rf{tanh}. 

Alternatively, following \cite{Carrasco:2015rva}, one can  make a change of variables $Z =\tanh
 {\Phi \over \sqrt {6\alpha}}$, where  $
\Phi \equiv \vp +i \vt \ .
$
Both procedures lead to the same canonically normalized $\vp$, but in the second case the variable $\vt$ describing the orthogonal direction  becomes canonical in the small  $\vt$ limit. 

In what follows, we will use both variables $\phi$ and $\varphi$; both of them are good for different purposes. 
The inflationary potential for the inflaton field $\vp$  at $\vt = 0$ is
\be
V= 3\alpha\, m^2 \tanh^2{ \vp\over \sqrt {6\alpha}}  = {m^{2}\over 2}\phi^{2}\ . 
\label{Vsimple}\ee
One can compare it with the first term in \rf{cosmo3} and confirm that $m^{2}$ is the mass squared of the inflaton field at the minimum $\vp = 0$. The full potential for the two variables is
\be
V= 3\alpha m^{2}\  {\cosh  \sqrt{2  \over 3\alpha} \vp - \cos  \sqrt{2  \over 3\alpha} \vartheta   \over \cosh  \sqrt{2  \over 3\alpha} \vp + \cos  \sqrt{2  \over 3\alpha} \vartheta  }
\cdot  \Big (\cos \sqrt{2  \over 3\alpha} \vartheta   \Big ) ^{-{3\alpha}} .
\label{Pot1}\ee

This potential 
has a minimum at $\vartheta=0$.
  At large values of $\vp$, where  $\tanh
 {\vp \over \sqrt {6\alpha}}$  approaches 1, the 
 potential in terms of $\vp$ and $\vt$ has a shift-symmetric dS valley of constant, $\vp$-independent width \cite{Carrasco:2015rva}.

One can better understand the structure of this potential along the inflationary flat direction by expanding it in powers of $\vartheta$ and $e^{-{\sqrt{2\over 3\alpha}\varphi}}$:
\be\label{expan}
V = V_{0}\left(1+ \vartheta^{2} - 4 e^{-{\sqrt{2\over 3\alpha}\varphi}}\Bigl(1+\vartheta^{2}\bigl(1-{1\over 3\alpha}\bigr)\Bigr)\right) +...
\ee
where $V_{0} = 3\alpha m^{2}$.  Thus, at $\vp \gg \sqrt\alpha$, the potential acquires a very simple asymptotic form, describing a dS valley, shift-symmetric with respect to the field $\vp$:
\be\label{expanshift}
V = V_{0} (1+\vartheta^{2}) \ ,
\ee
which helps to solve the problem of initial conditions for inflation in this class of models  \cite{Carrasco:2015rva, East:2015ggf}.

Now we are in  position to calculate masses and coupling constants along the stable inflationary trajectory $\vartheta = 0$. At large $\varphi$ one has, for $\vartheta = 0$:
\be\label{exp}
V =  V_{0}\left(1 - 4 e^{-{\sqrt{2\over 3\alpha}|\varphi|}}\right)\, ,
\ee
\be\label{slope}
{\partial_{\varphi} V}  = 4 V_{0} \sqrt{2\over 3\alpha}\, e^{-{\sqrt{2\over 3\alpha}\varphi}} \ ,\qquad {\partial_{\vartheta} V}  = 0\ ,
\ee
\be
m^{2}_{\varphi} = {\partial_{\varphi,\varphi} V} = -{8\over 3\alpha} V_{0} \, e^{-{\sqrt{2\over 3\alpha}\varphi}} = - {8} m^{2} \, e^{-{\sqrt{2\over 3\alpha}\varphi}} \ ,
\ee
\be\label{var}
m^{2}_{\vartheta} = {\partial_{\vartheta,\vartheta} V} = 2 V_{0} \Bigl(1 - 4 e^{-{\sqrt{2\over 3\alpha}\varphi}} \bigl(1-{1\over 3\alpha}\bigr) \Bigr)\, ,
\ee
Note that the slope of the potential and its curvature along the inflaton direction at large $\varphi$ are exponentially small, being suppressed by $e^{-{\sqrt{2\over 3\alpha}\varphi}}$, just as in the simple model studied in the previous section.  

In the theories with nilpotent field $S$ \cite{Ferrara:2014kva}, this field vanishes, so there is no need to calculate its mass. On the other hand, if one uses a standard unconstrained chiral superfield $S$, its mass squared  is given by
\be\label{ms2}
m^{2}_{s} = m^{2} \, \bigl(1-{\phi^{2}\over 6\alpha}\bigr)^{2} = m^{2} \, \bigl(1-\tanh^{2}{\varphi\over\sqrt {6 \alpha}}\bigr)^{2} = 16 m^{2} \, e^{-{\sqrt{2\over 3\alpha}\varphi}} \, ,
\ee
where the last expression is valid for $\vp \gg \sqrt\alpha$.

 Just as in the toy model discussed in the previous section,  one can easily see that all non-vanishing effective coupling constants describing strength of interaction of particles $\varphi$ and $\vartheta$, 
 are  suppressed by the same exponentially small factor $e^{-{\sqrt{2\over 3\alpha}\varphi}}$. This is true not only for the bosons, but also for their superpartners, which we did not present here explicitly. 
  
The same qualitative conclusions remain true for a more general class of the models with 
\be
W = S \, f(Z) \ ,
\ee  
where $f(Z)$ is an arbitrary holomorphic function, non-singular for $|Z| \leq 1$.  The plateau inflaton potential in these models is given by
\be
V = f^{2}(\tanh{ \vp\over \sqrt {6\alpha}}) \ .
\ee
In particular, superpotentials  $W \sim S \, Z^{n}$  lead to the T-models, with a T-shape potential symmetric with respect to the change $\vp\to -\vp$:
\be
V =V_{0}  \tanh^{2n}{ \vp\over \sqrt {6\alpha}} \ .
\ee
For $W \sim  {S Z\over 1+Z}$  one finds E-models, which represent $\alpha$-attractors with
\be
V = V_{0} (1- e^{-\sqrt{2/3\alpha}\,\varphi})^2 \ .
\ee
This class of models generalizes the potential of the Starobinsky model \cite{Starobinsky:1980te,whitt} and of its supergravity realizations proposed in \cite{Kallosh:2013xya,Farakos:2013cqa,Ferrara:2013rsa}, but in the context described above the asymptotic flatness of the potential  and the asymptotic freedom of the inflaton field at $\vp \gg \sqrt{6\alpha}$ are  protected by the geometric structure of the theory.

\section{Cosmological attractors and the rest of the world}\label{matter}

Here we will discuss a more advanced model with hyperbolic geometry, which describes  inflation and a matter multiplet $U$  with the \K\ potential \cite{Kallosh:2016ndd}
\be\label{KMATTER}
K= -{3  \alpha\over 2}  \log  \left[ {( 1- Z \overline{Z} )^2  \over  (1-Z^2) (1- \overline{Z}^2)}  \right] + S \overline{S} + U \overline{U}  ,  
\ee
and superpotential
\be
W =\sqrt{3\alpha}\, m S Z+ {M\over 2} U^{2} \ .
\ee
Representing $U = {x+i y\over \sqrt 2}$ and suppressing the imaginary component of the inflaton field $\varphi$ studied in the previous section, one finds the following expression for the potential of the canonically normalized fields $\vp$, $x$ and $y$: 
 \bea\label{Vsimple2}
V&=& e^{x^{2}+y^{2}\over 2}\Bigl(3\alpha m^2 \tanh^2{ \vp\over \sqrt {6\alpha}}   \\
&+&  
  {M^2\over 32} (x^2 + y^2) (16 + x^4 + 2 y^2 + y^4 + 2 x^2 (1 + y^2))\Bigr)\ .\nonumber 
\eea

Just as in the examples discussed above, the potential of  scalar fields $x$ and $y$ at large $\varphi$ has a flat direction along the inflationary trajectory  $x=y=0$, and the effective coupling constants of these fields to the inflaton field are suppressed by the exponential factor $e^{-{\sqrt{2\over 3\alpha}\varphi}}$. Meanwhile the squares of masses of these fields are equal to 
\bea
m^{2}_{{x,y}} &=& M^{2} + 3\alpha m^2 \tanh^2{ \vp\over \sqrt {6\alpha}} \nonumber\\ &= &M^{2} +V_{0}\left(1 - 4 e^{-{\sqrt{2\over 3\alpha} \varphi }}\right)+... 
\eea
In the large $\vp$ limit, the squares of these masses approach a constant value $M^{2} +V_{0}$.  

Interactions between the inflaton field and matter fields $U = {x+i y\over \sqrt 2}$ are described by the first term in \rf{Vsimple2}:
 \bea\label{Vint}
V_{\rm int}&=& 3\alpha m^2 e^{x^{2}+y^{2}\over 2}  \tanh^2{ \vp\over \sqrt {6\alpha}}\nonumber\\  &\approx&   3\alpha m^2 e^{x^{2}+y^{2}\over 2} \, (1-4 e^{-\sqrt{2\over 3\alpha}\varphi})
 \ . \eea
 Clearly, any coupling constant describing interaction of the inflaton field $\vp$ to the matter fields $x$ and $y$ will involve at least one derivative of with $V_{\rm int}$ with respect to $\vp$, and therefore it will be exponentially suppressed by $e^{-\sqrt{2\over 3\alpha}\varphi}$. 
 
 One can easily check that this result remains true if one replace the simple term $ {M\over 2} U^{2}$ in the superpotential by any non-singular function $F(Z,S,U)$. The choice of an arbitrary superpotential  may shift the direction of the inflationary trajectory in the moduli space if the fields $S$, $\vartheta$ and $U$ are not strongly stabilized, but it will not affect the fact of the exponential suppression of the coupling between the field $\vp$ and all other fields.
 
As for the stabilization, one can add to the \K\ potential the term $(S \overline{S})^{2} f_{1}(Z,\overline Z)$ stabilizing $S$ near $S = 0$, the term $S \overline{S} (Z-\overline Z)^{2} f_{2}(Z,\overline Z)$ for stabilization of $\vartheta$ near $\vartheta=0$, and the term $(S +\overline{S}) (Z-\overline Z)^{2} f_{3}(Z,\overline Z)$ for increasing the mass of the inflatino. With a proper choice of the functions $f_{i}(Z,\overline Z)$, these additions can bring our models very close to the theory of the orthogonal nilpotent fields \cite{Ferrara:2015tyn,Carrasco:2015iij,Dall'Agata:2015lek,Ferrara:2016een,Kallosh:2015pho,Dall'Agata:2016yof}.  

Alternatively, one may start directly with the theories of orthogonal nilpotent fields $S$ and $Z$ interacting with matter and simultaneously describing inflation, dark energy and supersymmetry breaking \cite{Carrasco:2015iij,Kallosh:2016ndd}. All our conclusions concerning the asymptotic freedom of the inflaton field and asymptotic flatness of the inflationary potential remain valid for these theories as well.

 \section{Quantum corrections to the inflaton potential}\label{CWsection}
 
 One may wonder whether quantum corrections can change the situation. For example, if the inflaton field gives mass $g\phi$ to some other fields, and the value of the field $\phi$ can be arbitrarily large, then the masses of these additional fields at super-Planckian values of the inflaton field become extremely large, which could lead to  large Coleman-Weinberg quantum corrections altering the shape of the original potential  \cite{Coleman:1973jx,Linde:1975sw,Weinberg:1976pe,Krive:1976sg,Linde:2005ht}.
 
In our case, the situation is different for two main reasons. First of all, the full range of the original field variables $\phi$ (or $Z$) is limited by the boundaries of the moduli space.  Secondly, even if quantum corrections are large, this may not matter much, because the cosmological predictions are mostly determined not by the full shape of the potential, but by its value $V_{0}$ near the boundary, which affects only the amplitude of perturbations of metric. But these changes can be rescaled back, by rescaling the original parameters. To be more precise, if quantum corrections change the shape of the potential far away from the boundary, it may affect the post-inflationary evolution and change the equation of state after inflation. This may lead to a slight change of the required number of e-foldings $N \sim 50 - 60$, which may slightly change the spectral index $n_{s}$. That is why we will briefly describe here the possible effects: They are not going to alter our main conclusions, but they may be useful for other reasons. 

As an example, let us consider the simple toy model \rf{cosmo2}, where the original inflaton field is confined in the region $-\sqrt{6\alpha} < \phi < \sqrt {6\alpha}$. As before, for definiteness we will consider $\phi > 0$, and assume that $\alpha = O(1)$ and $m^{2} \sim 10^{{-10}}$. One-loop Coleman-Weinberg corrections to the potential have the following general form \cite{Coleman:1973jx}:
\be\label{CW}
\Delta V(\vp) = \sum_{i}{m_{i}^{4}(\vp)\over 64\pi^{2}}\, \log{m^{2}(\vp)\over \Lambda^{2}} \ ,
\ee
where $\Lambda$ is a normalization constant and $m_{i}(\vp)$ are particle masses.  If all masses do not exceed  the inflaton mass, and the range of the fields is limited, $\phi < \sqrt {6\alpha}  = O(1)$, these corrections are suppressed by an extra factor $m^{2} \sim 10^{{-10}}$ as compared with the classical potential. 
Therefore we will concentrate on the most interesting (and dangerous) possibility that the field $\sigma$ is superheavy, $m^{2}_{\sigma} = M^{2} +  g^{2}\phi^{2} \gg m^{2}$, and therefore it might  give a significant  contributions to the inflaton potential.

The contribution $\Delta V_{\sigma}$ of the field $\sigma$ is given by
\be
\Delta V_{\sigma} =  {(g^{2}\phi^{2}+ M^{2})^{2}\over 64\pi^{2}}   \log{g^{2}\phi^{2}+ M^{2}\over \Lambda^2} \ .
\ee
Thus expression should be properly normalized: One can always add a constant vacuum energy, as well as the mass term $\sim \phi^{2}$ and the coupling term $\sim \phi^{4}$ without changing the structure of the theory. These terms can be chosen to protect the original properties of the theory at an arbitrary  normalization point. To properly relate all parameters to the post-inflationary phenomenology, it is convenient to chose the normalization point at $\phi = 0$, and add terms keeping zero cosmological constant (in our approximation), the original inflaton mass squared $m^{2} = \partial^{2}_{\phi}V$ at $\phi = 0$, and zero coupling $\lambda$ proportional to the fourth derivative of $V$ at $\phi = 0$  \cite{Linde:1975sw,Weinberg:1976pe,Krive:1976sg,Linde:2005ht}. 

A proper normalization of the cosmological constant is achieved by taking $\Lambda^2 = M^{2}$, which yields
\be\label{1}
\Delta V_{\sigma} =  {(g^{2}\phi^{2}+ M^{2})^{2}\over 64\pi^{2}}   \log{g^{2}\phi^{2}+ M^{2}\over M^2} \ .
\ee
Normalization $m^{2} = \partial^{2}_{\phi}V$ at $\phi = 0$ gives
\be\label{2}
\Delta V_{\sigma} =  {(g^{2}\phi^{2}+ M^{2})^{2}\over 64\pi^{2}}   \log{g^{2}\phi^{2}+ M^{2}\over M^2} -{g^{2}M^{2}\phi^{2}\over 64\pi^{2}}\ .
\ee
Finally, if one wants to have $V''''(0) = 0$, as in the theory \rf{cosmo2} in the tree approximation,  one ends up with
\be\label{3}
\Delta V_{\sigma}= {(g^{2}\phi^{2}+ M^{2})^{2}\over 64\pi^{2}}   \log\bigl(1+{g^{2}\phi^{2}\over M^2}\bigr) - {g^{2}\phi^{2}\over 128\pi^{2}}(2M^{2}+g^{2}\phi^{2}) \, . 
\ee
This potential has the  plateau functional form \rf{plateau} in terms of the canonically normalized inflaton field $\vp$, which means, in agreement with our general argument, that  quantum corrections do not change the asymptotic exponential flatness of the potential.   Similarly, quantum corrections to the potential do not change the  exponential smallness of the coupling constants at large values of the inflaton field. This suggests that quantum corrections to the kinetic term of the inflaton field $\varphi$ also remain negligibly small in the large field limit. In terms of the original field $\phi$, this means that quantum corrections do not alter the pole structure of the kinetic term in \rf{cosmo}, \rf{cosmo2}, which is consistent with the geometric origin of the pole discussed in section \ref{sugra}. 

To evaluate possible consequences of this result, let us assume that $g \phi \gg M^{2}$ at the boundary  $\phi  = \sqrt{6\alpha}$. In this case all three expressions \rf{1}, \rf{2} and \rf{3} have the same value, which does not depend on the renormalization of  $V''$ and $V''''$:
\be\label{qc}
\Delta V_{\sigma} \approx  {g^{4}\phi^{4}\over 64\pi^{2}}   \log{g^{2}\phi^{2}\over M^2}  \ .
\ee
In terms of the canonical field $\vp$, this is the plateau potential \rf{plateau}. The main effect of the additional term $\Delta V_{\sigma}$ that cannot be absorbed in the shift of the variable $\vp$ is the change of the height of the plateau $V_{0}$, which determines the amplitude of the scalar perturbations in this class of theories.  For $\alpha = O(1)$,  perturbations with the desirable amplitude are produced for $V_{0} \sim 10^{{-10}}$. 
In the tree approximation, one has $V_0 = V(\phi)|_{\phi = \sqrt {6 \alpha}} = 3\alpha m^{2}$, so one should have $m \sim 10^{-5}$. Let us assume that $g \phi \gg M^{2}$ at the boundary  $\phi  = \sqrt{6\alpha}$. In that case, quantum corrections \rf{qc} add to $3\alpha m^{2}$ the term
\be\label{CWtotalNumber}
\Delta V_{0} =  {9 \alpha^{2} g^{4}\over 16\pi^{2}}   \log {6 \alpha g^{2}\over M^2}    \ .
\ee 
To estimate the value of $\Delta V_{0}$, let us assume for simplicity that $\alpha = O(1)$ and $\log {6 \alpha g^{2}\over M^2} = O(10)$. This yields $\Delta V_{0} \sim g^{4}$. This term remains smaller than $10^{-10}$ and can be ignored  for $g \lesssim 3\times 10^{{-3}}$.

In other words, the simple expression for the inflaton potential \rf{ppot} remains valid unless the inflaton field interacts with superheavy fields, with  the masses $m_{\sigma} \sim g\phi$ approaching or exceeding the grand unification mass scale. But one may wonder whether one can use this fact constructively and study inflation in the theory where the potential is dominated by the quantum effects, $\Delta V \gg m^{2}\phi^{2}/2$.

Let us assume that $m \ll 10^{{-5}}$ and $m^{2}\phi^{2}/2 \ll 10^{-10}$, but  $\Delta V_{0} \sim 10^{{-10}}$. In that case one can ignore the tree level expression for the potential  as compared with the one-loop term $\Delta V_{0}$. The result obtained above implies that inflation in this theory produces perturbations with desirable amplitude if  $m_\sigma |_{\phi = \sqrt {6 \alpha}} = \sqrt{6 \alpha} g$  is close to the GUT mass scale. In other words, instead of the  fine-tuning of the inflaton mass $m \sim 10^{-5}$ in the tree level potential $m^{2}\phi^{2}/2 \ll 10^{-10}$ one can obtain inflationary perturbations of the same magnitude by taking into account quantum corrections $\Delta V_{0}$ due to particles $\sigma$ with the GUT mass scale $m_{\sigma} \sim 10^{-2} - 10^{{-3}}$. 

An opposite limit $ g^{2}\phi^{2} \ll M^{2}$ is equally interesting. In this case \rf{3} yields
\be\label{small}
\Delta V_{\sigma} \approx  
{g^{6}\phi^{6} \over 192 \pi^{2} M^{2}}  \ .
\ee
In this case one can have $\Delta V_{0} \sim 10^{{-10}}$ and $m^{2}\phi^{2}/2 \ll 10^{-10}$ for $M > 10^{{-2}}$ and $g < 10^{{-3}}$. Notice that in this case the post-inflationary potential instead of being proportional to $\phi^{2}$ is proportional to $\phi^{6}$.

In this class of models, the spectral index $n_{s} = 1-2/N$  for $\alpha = O(1)$ does not depend much on the choice of the potential in the leading approximation in $1/N$, where $N \sim 50 - 60 $ is the  number of e-foldings. But the required  number of e-foldings $N$ does depend on the post-inflationary evolution and therefore on the choice of the potential at $\phi \ll \sqrt{6\alpha}$. This effect is rather small,  higher order in $1/N$, but it is potentially observable \cite{Bezrukov:2011gp}. 
 
 Suppose e.g. that reheating is very inefficient, which is often the case in supergravity where the inflaton field may belong to the hidden sector. After inflation, the  field  $\phi$ oscillates and immediately becomes much smaller than  $ \sqrt{6\alpha}$, so it becomes almost exactly canonically normalized.  In the theory with a quadratic potential, an effective equation of state of the oscillating inflaton field after inflation but before the end of reheating is  $w \approx 0$.  In the theory with a potential $\sim \phi^{4}$  \rf{qc}, the oscillating field $\phi \lesssim 1$ after inflation has an effective equation of state $w \approx 1/3$. This may lead to an increase of $N$ in the models with $w \approx 1/3$ by about $\Delta N  \sim 5$. This results in a slight increase of $n_{s}$ by $\Delta n_{s} \sim 2\Delta N/N^{2} \sim 4\times 10^{{-3}}$. Meanwhile  equation of state of the oscillating field $\phi$  with the potential \rf{small} is $w \approx 1/2$, which may lead to further increase of $n_{s}$. These changes are small, but they may help these models to match the observational data even better \cite{Ueno:2016dim,Eshaghi:2016kne}.  If more precise observational constraints on $n_{s}$ become available, it may allow us to distinguish between different classes of the cosmological attractors.

\section{Conclusions}

It this paper we revealed a simple and universal property of cosmological  $\alpha$-attractors, where inflation is determined by physical processes near a pole of the kinetic term as shown, e.g., in the Lagrangian \rf{cosmo}. In all such models, it is convenient to formulate the theory in terms of the  original field variables $\phi$, which properly describe the underlying geometry of the moduli space \cite{Kallosh:2015zsa}. Upon a transformation  to canonical variables $\vp$, which are more suitable for the description of inflation, the potential of the inflaton field becomes exponentially stretched and acquires asymptotic form $V(\vp)$ practically independent of the functional form of the potential $V(\phi)$ in the original geometric variables. We found that this effect typically implies that coupling constants describing the strength of interaction of the inflaton field with other fields become exponentially small at large $\vp$, being suppressed by $e^{-{\sqrt{2\over 3\alpha}\varphi}}$. Meanwhile masses of all fields interacting with the inflaton field tend to approach constant values at large $\vp$. They deviate from these asymptotic values only by terms suppressed by $e^{-{\sqrt{2\over 3\alpha}\varphi}}$.

It is too early to make any statements about the fundamental significance of these results beyond phenomenological supergravity, but we believe that  the existence of a broad class of cosmological attractors with these properties is quite intriguing.

\subsection*{\bf Acknowledgements} 
 We are grateful to J. J. Carrasco, S. Ferrara, A. Van Proeyen, D. Roest, J. Thaler and T. Wrase for  enlightening  discussions and collaboration on related projects.   The work of RK  and AL is supported by the SITP, and by the NSF Grant PHY-1316699.  AL is also supported by the Templeton foundation grant `Inflation, the Multiverse, and Holography'.

\end{document}